\documentstyle[preprint,floats,epsf,tighten,aps,eqsecnum]{revtex}
\global\arraycolsep=2pt 
\input{epsf}
\makeatletter
\def\fmslash{\@ifnextchar[{\fmsl@sh}{\fmsl@sh[0mu]}}
\def\fmsl@sh[#1]#2{%
  \mathchoice
    {\@fmsl@sh\displaystyle{#1}{#2}}%
    {\@fmsl@sh\textstyle{#1}{#2}}%
    {\@fmsl@sh\scriptstyle{#1}{#2}}%
    {\@fmsl@sh\scriptscriptstyle{#1}{#2}}}
\def\@fmsl@sh#1#2#3{\m@th\ooalign{$\hfil#1\mkern#2/\hfil$\crcr$#1#3$}}

\makeatother
\newcommand{\mst}[2]{\mbox{\raisebox{-1mm}{$\,\stackrel{#1}{\scriptstyle
#2}\,$}}}

\def\back#1{\raise1.5ex\hbox{$\leftarrow$}\mkern-16.5mu #1}

\def\half{{\textstyle{1\over 2}}}
\def\lqcd{\Lambda_{\rm QCD}}
\def\OMIT#1{{}}
\def\nslash{n\!\!\!\slash}

\newcommand{\bn}{\bar n}
\def\model{{\rm{mod}}}
\def\GeV{{\rm GeV}}

\begin{document}

\draft

{\tighten

\preprint{\vbox{\hbox{CERN-TH 2001-027}
                \hbox{TTP00-27}
                \hbox{UCSD/PTH 00-33}
                \hbox{UTPT--00-14}
                \hbox{hep-ph/0102089}}}

\title{Light-cone distribution functions for $B$ decays \\
at subleading order in $1/m_b$}

\author{Christian W. Bauer$^{a}$, Michael Luke$^{b}$ 
and Thomas Mannel$^{c,d}$}

\address{ \vbox{\vskip 0.truecm}
$^{a}$Department of Physics, University of California, San
Diego \\ 9500 Gilman Drive, La Jolla CA 92093\\
$^{b}$Department of Physics, University of Toronto \\
60 St.\ George Street, Toronto, Ontario, Canada M5S 1A7\\
$^{c}$Institut f\"{u}r Theoretische Teilchenphysik,
Universit\"{a}t Karlsruhe \\ D--76128 Karlsruhe, Germany\\
$^{d}$TH Division, CERN, CH-1211 Geneva 23, Switzerland
}

\maketitle

\begin{abstract}%
We construct the higher twist structure functions that describe inclusive $b$
hadron decays in regions of phase space where the hadronic decay products carry
high energy but have low invariant mass. We show that, for $B$ meson decays,
there are four new non-vanishing matrix elements of non-local operators. We show
that to subleading twist these decays are parametrized in terms of four
functions. We compute the tree-level matching for a general heavy-to-light
current and apply it to $\bar B \to X_s \gamma$. Using a simple model for these
functions we estimate the subleading twist contributions to this decay.
\end{abstract}

}
\newpage
\tighten

\section{Introduction}
Inclusive decays of $B$ mesons can be analyzed via an operator product expansion
(OPE) \cite{OPE} as a power series in $\lqcd/m_b$. At leading order the OPE
reproduces the parton model results, while non-perturbative physics is
parametrized by matrix elements of higher dimensional operators and suppressed by
powers of the heavy quark mass.

In practice, higher dimensional operators in the OPE are only suppressed for
sufficiently inclusive observables.   In the resonance regime, where the
invariant mass of the final hadronic state is restricted to be $\mst{<}{\sim}
\lqcd^2$, the decay is no longer inclusive and it is not surprising that the OPE
completely breaks down.  However, the OPE includes terms suppressed by
$E_X\lqcd/m_X^2$ (where $E_X$ and $m_X$ are the energy and invariant mass of the
final hadronic state) which are suppressed over most of phase space, but are
$O(1)$ in the region of high energy, low invariant mass hadronic states,
\begin{equation}\label{shapefnregime}
E_X\sim m_b,\ m_X^2\sim \lqcd m_b\gg\lqcd^2\,.
\end{equation}
In this ``shape function region" the OPE breaks down, even though it is far from
the resonance regime. Thus an inclusive description is still appropriate, but the
expansion parameter in the OPE has to be modified. It has been shown that the
most singular terms in the OPE may be resummed into a non-local
operator\cite{shapefunct}
\begin{eqnarray}\label{O(y)}
O_0(\omega) = \bar h_v \delta(\omega +in \cdot \hat D) h_v\,,
\end{eqnarray}
where $n^\mu$ is a light-like vector in the direction of the final hadrons, $h_v$
is a HQET heavy quark field and we denote variables normalized to $m_b$ by a
hat: $\hat{D}^\mu\equiv D^\mu/m_b$. The matrix element of this operator in a $B$
meson is the light-cone structure function of the meson,
\begin{eqnarray}\label{distfn}
f(\omega)={1\over 2m_B}\langle B|O_0(\omega)|B\rangle\,.
\end{eqnarray}
The rate in the shape function region is determined by $f(\omega)$. However,
since it is a non-perturbative function, $f(\omega)$ cannot be calculated
analytically, and the rate in the shape function region is model-dependent even
at leading order in $\lqcd/m_b$.

Unfortunately, for  $\bar B\rightarrow X_u\ell\bar\nu$ decay (and to a lesser
extent $\bar B\rightarrow X_s\gamma$), the experimental cuts which must be
imposed on the phase space to eliminate large backgrounds from $b\rightarrow c$
decay typically put the decay into the shape function region, introducing large
 model-dependence in the predicted rate.  For semileptonic $b\rightarrow u$
decay, this is the case for cuts on either the charged lepton energy or the
hadronic invariant mass\cite{invtmass}, and this model dependence is the major
theoretical stumbling block to a precise determination of the CKM matrix element
$|V_{ub}|$ from inclusive decays.

However, since the distribution function (\ref{distfn}) determines the shape of
the photon spectrum in $\bar B\rightarrow X_s\gamma$ as well as the charged
lepton or hadronic invariant mass spectrum in $\bar B\rightarrow X_u\ell\bar\nu$,
it was suggested a number of years ago\cite{mn} that $f(y)$ could be measured in
$\bar B\rightarrow X_s\gamma$, and then used to extract $|V_{ub}|$ from
semileptonic decay.\footnote{Another solution is to consider a kinematic cut
which does not put the final state into the shape function region, such as a cut
on the lepton invariant mass\cite{bll99}.}  The perturbative corrections to the
relation between these processes have been intensely studied in recent
years\cite{sudakovs}.

In addition to the radiative corrections, however, there are non-perturbative
corrections to the relation between the photon spectrum in $\bar B\rightarrow X_s
\gamma$ and the lepton or hadronic invariant mass spectra in $\bar B\rightarrow
X_u\ell\bar\nu$. The light-cone distribution function (\ref{distfn}) only resums
the most singular terms of the OPE in the shape function region. There are
corrections to this from less singular terms suppressed by $\lqcd/m_b$, analogous
to higher-twist corrections to deep inelastic scattering \cite{efp83}.  These
corrections are important for a precision measurement of $|V_{ub}|$ via this
method, but have not yet been studied.

In this paper we discuss the subleading corrections to heavy-light $B$ decay in
the shape function region by performing a twist expansion rather than the usual
OPE in terms of local operators\footnote{ As discussed in \cite{bfl00}, there are
actually two stages of matching: at $\mu=m_b$ QCD is matched onto an intermediate
theory with collinear and soft degrees of freedom, while at a lower scale the
non-local OPE is performed.  Since we are not concerned with summing Sudakov
logarithms in this paper, we may neglect the intermediate theory.}. At leading
order in the twist expansion we reproduce the known results, while at subleading
order we find four new non-local operators relevant for $B$ decays.  We compute
the tree-level matching for a general heavy-light current and apply it to $\bar
B\rightarrow X_s\gamma$, reserving a discussion of $\bar B\rightarrow
X_u\ell\bar\nu$ decay for a future work\cite{blmprog}.  We use a simple model for
these functions to estimate the subleading twist contribution to $B\rightarrow
X_s\gamma$.

\section{Kinematics}
\label{kinematics}

We consider a general heavy to light transition (radiative or semileptonic),
proceeding via the current
\begin{equation}
j(x) = \bar{q}(x) \Gamma b (x),
\end{equation}
where $q(x)$ is a massless quark field and $\Gamma$ is an arbitrary Dirac matrix.
The decay rate is related to the imaginary part of the $T$-product of two
heavy-light currents:
\begin{equation}
d\Gamma \sim -2 \, {\rm Im}\, \langle B|{\cal T}|B  \rangle\,,
\end{equation}
where
\begin{equation} \label{T1}
{\cal T}(q)  = \int\! d^4 x \, T \left[ j(0) j^\dagger (x) \right] e^{iqx}\,,
\end{equation}
and $q$ is the momentum transfer.

\begin{figure}[t]
\centerline{\epsfysize=3.5truecm \epsfbox{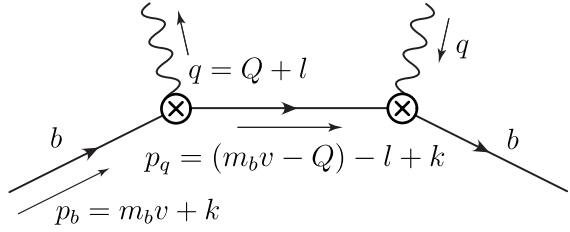} }
\caption[]{Kinematics for a general heavy to light transition.}
\label{kinematicsfig}
\end{figure}

The kinematics for this process are shown in Fig.\ \ref{kinematicsfig}.
As usual, the heavy quark momentum is split into a large and a residual piece,
$p_b^\mu= m_b v^\mu +k^\mu$.  In the shape function region the final hadronic
state has large energy but small invariant mass, and so its momentum lies close
to the light-cone.  We introduce a light-like vector $n^\mu$ to define the
expansion of the jet about the light-cone, and the velocity of the heavy quark
then defines a second light-like vector $\bar{n} = 2v-n$. These vectors satisfy
\begin{equation}
n^2 = 0 = \bar{n}^2, \quad v\cdot n = v\cdot \bar{n} = 1, \quad n\cdot
\bar{n} = 2.
\end{equation}
In the frame in which the $b$ quark is at rest and the emitted hadrons move in
the $+z$ direction, these vectors are given by $n^\mu=(1,0,0,1)$, $\bar
n^\mu=(1,0,0,-1)$ and $v^\mu=(1,0,0,0)$.  We can decompose the metric according
to
\begin{equation}
g^{\mu \nu} = \frac{1}{2} \left(n^\mu \bar{n}^\nu + n^\nu \bar{n}^\mu \right)
              + g^{\mu \nu}_\perp\,,
\end{equation}
which defines $g^{\mu \nu}_\perp$.

In the shape function region, the momenta scale as
\begin{eqnarray} \label{qscaling}
(m_b v - q)\cdot \bar n &=& m_b - q\cdot \bar n \sim  O(m_b), \nonumber \\
(m_b v - q)\cdot n &=& m_b - q\cdot n \sim O(\lqcd), \\
k^\mu&\sim&O(\lqcd).\nonumber
\end{eqnarray}
It is therefore convenient to split the momentum $q^\mu$ into large and small
components,
\begin{equation}
q^\mu\equiv Q^\mu+\ell^\mu\,,
\end{equation}
where
\begin{equation}
Q^\mu=\half (m_b\bar n^\mu+q\cdot\bar n n^\mu),\ \ \
\ell^\mu=-\half(m_b-q\cdot n)\bar n^\mu
\end{equation}
are $O(m_b)$ and $O(\lqcd)$, respectively.
The momentum of the light hadronic decay products is
\begin{equation}
p_q^\mu=m_b v^\mu-Q^\mu-\ell^\mu+k^\mu=\half (m_b-q\cdot \bar n) n^\mu
+\half(m_b-q\cdot n)\bar n^\mu+k^\mu\,,
\end{equation}
and so in the shape function region (\ref{qscaling}) we have $p_q^2={\cal
O}(\lqcd m_b)\ll m_b^2$.

\section{Matching}

\subsection{Leading Order}

The expansion in the shape function region differs from the usual $1/m_b$
expansion because of the additional small parameter $m_b-q\cdot n$; in the usual
OPE terms of order
$$
{k^\mu\over m_b-q\cdot n}
$$
are treated as subleading, whereas they are $O(1)$ in the shape function region.
Thus, instead of resumming terms of this form to all orders, we perform an OPE in
powers of $\lqcd/m_b$, but using the scaling (\ref{qscaling}). This is analogous
to the twist expansion in DIS.

Expanding the light-quark propagator shown in Fig.\ \ref{kinematicsfig} in powers
of $\lqcd/m_b$ gives
\begin{equation}\label{expanddenom}
{m_b \fmslash{v}+\fmslash{k}-\fmslash{Q}-\fmslash{l}\over
\left(m_b v+k-Q-l+i\epsilon\right)^2}={1\over 2}\,{\fmslash{n}\over m_b-q\cdot n+k\cdot n
+i\epsilon}+
O\left({\lqcd\over m_b}\right).
\end{equation}
Since $k\cdot n$ and $m_b-q\cdot n$ are of the same order, this term cannot be
expanded in powers of $k\cdot n$, and so cannot be matched onto a finite set of
local operators.

Instead, consider the set of operators
\begin{equation}
\tilde O_0(t)=\bar h_v(0) E(0,t) h_v(t)\,,
\end{equation}
where we use the shorthand notation
\begin{equation}
\psi(t)\equiv \psi\left({n t}\over m_b\right),
\end{equation}
where $t$ is dimensionless, to denote fields on the light-cone defined by
$n^\mu$.  The path-ordered exponential
\begin{equation}
E(t_1,t_2)=P\exp\left(-i\int_{t_1}^{t_2}
n\cdot  \hat A(t^\prime)\,dt^\prime \right)
\end{equation}
is required to make the operator $\tilde O_0(t)$ gauge invariant.  The operators
$O_0(\omega)$ are defined in terms of $\tilde O_0(t)$ by the linear combination
\begin{equation}\label{ftO0}
O_0(\omega) = {1\over 2\pi}\int_{-\infty}^{\infty} dt\,e^{-i\omega t} O_0(t)
=\bar h_v \delta(\omega+in\cdot \hat D) h_v\,,
\end{equation}
and have the required form for the imaginary part of the leading term in the
heavy quark expansion (\ref{expanddenom}).   The matrix elements of $O_0(\omega)$
define the light-cone distribution function of the $b$ quark in a $B$ meson
\begin{equation} \label{shapef}
f(\omega)\equiv {1\over 2 m_B}
\langle B |\bar h_v\delta(\omega+in \cdot \hat D)h_v|B\rangle\,.
\end{equation}
Expanding $f(\omega)$ in powers of $in\cdot \hat D$ gives the series of
increasingly singular terms
\begin{equation}\label{leadexp}
f(\omega) = \delta(\omega)
              - \frac{\lambda_1}{6m_b^2} \delta '' (\omega) -
                \frac{\rho_1}{18m_b^3} \delta ''' (\omega) + \cdots \nonumber\\
\end{equation}
where
\begin{eqnarray}
{1\over 2m_B}\langle B|\bar h_v(iD_\alpha)(iD_\beta) h_v | B\rangle&\equiv&
{1\over 3}(g_{\alpha\beta}-v_\alpha v_\beta)\lambda_1\\
{1\over 2m_B}\langle B|
(iD_\alpha)(iD_\mu)(iD_\beta)|B\rangle&\equiv&{1\over 3}(g_{\alpha\beta}-v_\alpha
v_\beta)v_\mu\rho_1.
\end{eqnarray}
For a general $b$ hadron decay there is an additional parity-odd operator at
leading twist
\begin{equation}\label{ftP0}
\tilde P_0^\alpha(t) =\bar h_v(0)\gamma^\alpha\gamma_5 E(0,t) h_v(t) \,,
\end{equation}
whith $P_0^\alpha(\omega)$ defined analogously to (\ref{ftO0}). This operator is
not relevant for $B$ meson decays since its matrix element vanishes,  but it
gives a spin dependent contribution to $\Lambda_b$ decay.  A general Dirac
structure between heavy quark fields may be expressed in terms of these four
independent matrices via the  projection formula
\begin{equation}\label{projform}
P_+ \Gamma P_+ = \frac{1}{2} P_+ {\rm Tr}( P_+ \Gamma ) -
                  \frac{1}{2} s_\mu {\rm Tr} (s^\mu \Gamma) \, ,
\end{equation}
where
\begin{equation}
s_\mu \equiv P_+ \gamma_\mu \gamma_5 P_+\,,
\end{equation}
and $P_+ \equiv \frac12(1+\fmslash{v})$.

Since the OPE is performed over a a continuously infinite set of operators
labeled by $\omega$, the heavy quark expansion in the shape function region is,
like in DIS, a convolution over a single parameter which may be interpreted as
the light-cone momentum fraction of the heavy quark:
\begin{equation}\label{leadingconv}
{\rm Im}\,{\cal T}(q)=-{1\over 2m_b}\int_{-\infty}^\infty\! d\omega\, \left[C_0(\omega)
O_0(\omega)+ C_{5,0}^\alpha(\omega) P_{0,\alpha}(\omega)+O\left({\lqcd\over
m_b}\right)\right]\,,
\end{equation}
where the $C_i$'s are perturbatively calculable short distance coefficients. The
tree level matching conditions are easily obtained from (\ref{expanddenom}) and
(\ref{projform}):
\begin{eqnarray}
C_0(v,q, \omega) &=& \frac{\pi}{2}{\rm Tr } \left(P_+ \Gamma \nslash            
     \overline{\Gamma} \right) \delta(1 - n \cdot \hat q - \omega)              
   \\ C_{5, 0}^{\alpha}(v,q, \omega) &=& - \frac{\pi}{2}{\rm Tr }
\left(s^\alpha\Gamma \nslash \overline{\Gamma} \right)\delta(1 -
n \cdot \hat q -\omega ) \,.
\end{eqnarray}

\subsection{Subleading Order}

Expanding (\ref{expanddenom}) to subleading order in $1/m_b$, we will in general
match onto non-local objects of the form\cite{efp83}
\begin{equation}
\tilde O_\Gamma^{\mu_1 \dots \mu_{n-1}}(t_1, \dots , t_n) =
\bar h_v(0) \Gamma \big[iD^{\mu_1}(t_1) \dots iD^{\mu_{n-1}}(t_{n-1})\big] h_v(t_n)\,,
\end{equation}
where $\Gamma=1$ or $\gamma_\alpha \gamma_5$ and
\begin{equation}
D_\mu(t)\equiv \partial_\mu-ig  A_\mu^a(t) T^a
\end{equation}
is the usual covariant derivative, acting at a light-cone coordinate.

The most general case involves operators in which every field and derivative is
evaluated at a different light-cone coordinate. However for heavy quark decays to
subleading order, we find that at maximum only two light-cone coordinates enter.
The complete set of operators required is:
\begin{eqnarray}\label{nonlocalops}
\tilde O_1^\mu(t)&=& \bar
h_v(0)\left(-i\back{D}^\mu(0)E(0,t)+E(0,t)iD^\mu(t)\right) h_v(t)
\nonumber\\
\tilde P_{1,\alpha}^\mu(t)&=& \bar
h_v(0)\left(-i\back{D}^\mu(0)E(0,t)+E(0,t)iD^\mu(t)\right)
\gamma_\alpha\gamma_5 h_v(t) \nonumber \\
\tilde O_2^\mu(t)&=& i \bar
h_v(0)\left(-i\back{D}^\mu(0)E(0,t)-E(0,t)iD^\mu(t)\right) h_v(t)
\nonumber \\
\tilde P_{2,\alpha}^\mu(t)&=& i \bar
h_v(0)\left(-i\back{D}^\mu(0)E(0,t)-E(0,t)iD^\mu(t)\right)
\gamma_\alpha\gamma_5 h_v(t) \\
\tilde
O_3^{\mu\nu}(t_1,t_2)&=& \bar h_v(0) E(0,t_1)
\left\{iD_\perp^\mu(t_1), iD_\perp^\nu(t_1)\right\} E(t_1,t_2)h_v(t_2)
\nonumber\\
\tilde P_{3,\alpha}^{\mu\nu}(t_1,t_2)&=& \bar h_v(0)
E(0,t_1)\left\{iD_\perp^\mu(t_1),
iD_\perp^\nu(t_1)\right\}E(t_1,t_2)\gamma_\alpha\gamma_5 h_v(t_2)
\nonumber\\
\tilde O_4^{\mu\nu}(t_1,t_2)&=& g\bar h_v(0) E(0,t_1)
G^{\mu\nu}_\perp(t_1) E(t_1,t_2)h_v(t_2) \nonumber\\
\tilde
P_{4,\alpha}^{\mu\nu}(t_1,t_2)&=& g\bar h_v(0)
E(0,t_1)G^{\mu\nu}_\perp(t_1)E(t_1,t_2) \gamma_\alpha\gamma_5
h_v(t_2)\,, \nonumber
\end{eqnarray}
where $D_\perp^\mu= g^{\mu \nu}_\perp D_\nu$ and $g \,G^{\mu \nu}_\perp = i
[(iD^\mu_\perp),(iD^\nu_\perp)]$ is the gluon field strength.

\begin{figure}[tbh]
\centerline{\epsfxsize=14cm \epsfbox{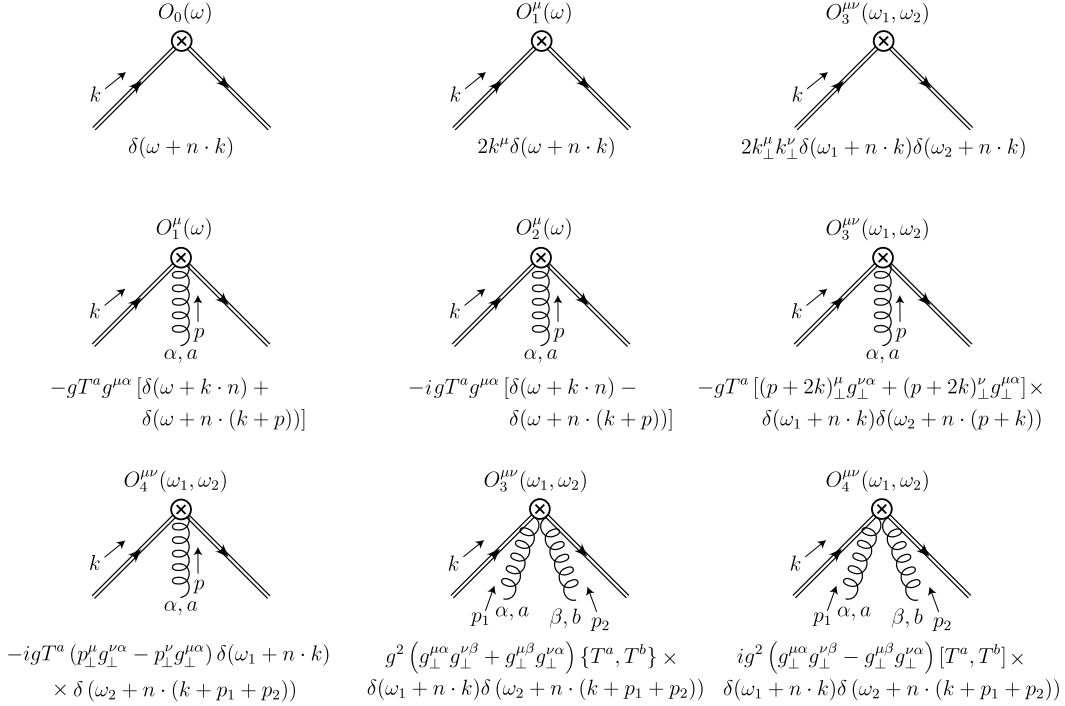} }
\caption[]{Feynman rules for non-local operators $O_0-O_4$ in $n\cdot A=0$
gauge.
The Feynman rules for $P_0-P_4$ are identical except for the Dirac structure.}
\label{nonlocalrules}
\end{figure}

The Fourier transformed operators $O_i$ are defined as
\begin{eqnarray}\label{nonlocalfteven}
O_1^\mu(\omega)&=&  {1\over 2\pi}\int_{-\infty}^{\infty} dt\,e^{-i\omega t}
\tilde O_1^\mu(t) =
\bar h_v \left\{i D^\mu,\delta(in \cdot \hat D + \omega)\right\} h_v
\nonumber\\
O_2^\mu(\omega)&=&{1\over 2\pi}\int_{-\infty}^{\infty} dt\,e^{-i\omega t}
\tilde O_2^\mu(t) =
i\bar h_v \left[i D^\mu,\delta(in \cdot \hat D + \omega)\right] h_v
\nonumber\\
O_3^{\mu\nu}(\omega_1,\omega_2) &=& \left({1\over 2\pi}\right)^2
\int_{-\infty}^{\infty} dt_1 \, dt_2\,
e^{-i\omega_1 t_1}  e^{-i(\omega_1 - \omega_2) t_2 }
\tilde O_3^{\mu\nu}(t_1,t_2) \\ \nonumber
&=&\bar h_v \delta(in \cdot \hat{D}+\omega_2)\left\{iD^\mu_\perp,
iD^\nu_\perp\right\}\delta(i n \cdot \hat{D}+\omega_1) h_v
\\ \nonumber
O_4^{\mu\nu}(\omega_1,\omega_2) &=&  \left({1\over 2\pi} \right)
\int_{-\infty}^{\infty} dt_1 \, dt_2\,
e^{-i\omega_1 t_1}  e^{-i(\omega_1 - \omega_2) t_2 }
\tilde O_4^{\mu\nu}(t_1,t_2) \nonumber \\
&=& g \bar h_v \delta(in \cdot \hat{D}+ \omega_2)G_\perp^{\mu\nu}\delta(i n
\cdot \hat{D}+\omega_1) h_v \, .
\nonumber
\end{eqnarray}
Similarly the Fourier transforms  of the $\tilde P_i$'s are
\begin{eqnarray}\label{nonlocalftodd}
P_{1,\alpha}^\mu(\omega)&=&
\bar h_v \left\{i D^\mu,\delta(in \cdot \hat D + \omega)\right\}
\gamma_\alpha\gamma_5 h_v
\nonumber\\
P_{2,\alpha}^\mu(\omega)&=&
i\bar h_v \left[i D^\mu,\delta(in \cdot \hat D + \omega)\right] \gamma_\alpha\gamma_5 h_v
\\
P_{3,\alpha}^{\mu\nu}(\omega_1,\omega_2) &=&
\bar h_v \delta(in \cdot \hat{D}+\omega_2)\left\{iD^\mu_\perp,
iD^\nu_\perp\right\}\delta(i n \cdot \hat{D}+\omega_1)\gamma_\alpha\gamma_5  h_v
\nonumber
\\ \nonumber
P_{4,\alpha}^{\mu\nu}(\omega_1,\omega_2) &=&
g \bar h_v \delta(in \cdot \hat{D}+ \omega_2)G_\perp^{\mu\nu}\delta(i n
\cdot \hat{D}+\omega_1)\gamma_\alpha\gamma_5  h_v.
\nonumber
\end{eqnarray}
The Feynman rules for the operators $O_0 - O_4$ in $n\cdot A=0$ gauge are shown
in Fig.~\ref{nonlocalrules}.

Finally, at subleading order there are also contributions from the time-ordered
products of $O_0(\omega)$ with the subleading terms in the HQET Lagrangian,
\begin{equation}
{\cal O}_{1/m} (y) = \bar{h}_v (y) (iD)^2 h_v (y)+{g\over 2} \bar h_v(y)
\sigma_{\mu\nu} G^{\mu\nu} h_v(y)\,.
\end{equation}
This  yields another two operators
\begin{eqnarray}\label{T-ops}
O_T (\omega) &=& i \int\! d^4y \, \frac{1}{2 \pi} \int\! dt \,
  e^{-i\omega t} T \left( \bar{h}_v (0) h_v (t) {\cal O}_{1/m} (y)
  \right) \\ \nonumber
P_{T,\alpha} (\omega) &=& i \int\! d^4y \,
  \frac{1}{2 \pi} \int\! dt \, e^{-i\omega t} T \left( \bar{h}_v (0)
  \gamma_\alpha \gamma_5 h_v (t) {\cal O}_{1/m} (y) \right)\,.
\end{eqnarray}
At subleading order the nonlocal OPE in (\ref{leadingconv}) is
\begin{eqnarray}\label{forward}
-2m_b \,{\rm Im}\, {\cal T}(q) &=&  \vphantom{{\lqcd^2\over m_b^2}}
          \int\! d\omega\,\left(
                    C_0(v, q,\omega) O_0 (\omega)
                  + C_{5,0}^\alpha(v,q,\omega)P_{0, \alpha} (\omega)
                  \right)\nonumber\\
&&\qquad+\frac{1}{2 m_b} \sum_{i=1,2}\int\! d\omega \, \left(
                    C_i^\mu (v,q,\omega) O_{i,\mu} (\omega)
                  + C_{5,i}^{\alpha,\mu} (v,q,\omega) P_{i, \alpha, \mu} (\omega)
                  \right)\nonumber\\
&&\qquad+ \frac{1}{2 m_b} \sum_{i=3,4}\int\! d\omega_1 d\omega_2 \, \left(
                    C_i^{\mu\nu}(v, q,\omega_1, \omega_2) O_{i,\mu\nu}
(\omega_1,\omega_2)\right.\\
&&\qquad\qquad\qquad\qquad\left.+ C_{5,i}^{\alpha,\mu\nu}(v,q,\omega_1, \omega_2)
                            P_{i,\alpha, \mu\nu}(\omega_1,\omega_2)
                  \right)\nonumber\\
&&\qquad+ \frac{1}{2 m_b} \int\! d\omega \, \left( C_T  (v, q,\omega) O_T (\omega)
                         + C_{5,T}^\alpha  (v,q,\omega) P_{T,\alpha} (\omega) \right)
\nonumber \\
&&\qquad+O\left({\lqcd^2\over m_b^2}\right)\,.\nonumber
\end{eqnarray}

The matching at subleading order onto the operators (\ref{nonlocalops}) is
performed by computing the zero, one and two gluon matrix elements of (\ref{T1})
in full QCD and comparing this to the operators in (\ref{nonlocalfteven}) and
(\ref{nonlocalftodd}). Note that this includes terms from the expansion of the
$b$ quark field,
\begin{equation}
b=\left(1+{i\fmslash{D}\over{2m_b}}+\dots \right) h_v.
\end{equation}
At tree level, we find
\begin{eqnarray} \label{coeffeven}
C_1^\mu (v, q,\omega) &=& \frac{\pi}{4} {\rm Tr } \left( \{P_+,\gamma^\mu\}
\Gamma \nslash \overline{\Gamma} \right) \delta(1 - n \cdot \hat q -\omega)
\nonumber \\
C_2^\mu (v, q,\omega) &=&  (-i) \frac{\pi}{4} {\rm Tr } \left( [P_+,\gamma^\mu   ]
\Gamma \nslash \overline{\Gamma} \right) \delta (1 - n \cdot \hat q - \omega) \nonumber\\
C_3^{\mu\nu} (v, q,\omega_1,\omega_2)
&=& - \frac{\pi}{2} g_\perp^{\mu\nu}
\frac{1}{m_b-\bar n \cdot q }
{\rm Tr } \left( P_+ \Gamma \nslash \overline{\Gamma} \right)\\
&&\qquad\times
\left(\frac{\delta(1 - n \cdot \hat q - \omega_1) - \delta(1 - n \cdot \hat q - \omega_2)}
            {\omega_1 - \omega_2} \right)
\nonumber  \\
C_4^{\mu\nu}  (v, q,\omega_1,\omega_2) &=& i \, \frac{\pi}{2}
\frac{1}{m_b-\bar n \cdot q }
{\rm Tr } \left( P_+ \Gamma \nslash (-i \sigma_{\!\!\perp}^{\mu\nu}) \overline{\Gamma} \right)
\nonumber\\
&&\qquad\times\left(\frac{\delta(1 - n \cdot \hat q - \omega_1) - \delta(1 - n \cdot \hat q -
\omega_2)}
            {\omega_1 - \omega_2} \right)
\nonumber \\ \nonumber
C_T (n \cdot q,\omega)  &=& C_0 (n \cdot q,\omega) \,,
\end{eqnarray}
and for the corresponding spin dependent operators
\begin{eqnarray} \label{coeffodd}
C_{5,1}^{\alpha,\mu} (v, q,\omega) &=& -\frac{\pi}{4}
{\rm Tr } \left( \{s^\alpha, \gamma^\mu \} \Gamma \nslash \overline{\Gamma} \right)
\delta(1 - n \cdot \hat q - \omega)
\nonumber \\
C_{5,2}^{\alpha,\mu} (v, q,\omega) &=&  i\,\frac{\pi}{4}
{\rm Tr } \left([s^\alpha, \gamma^\mu] \Gamma \nslash \overline{\Gamma} \right)
\delta(1 - n \cdot \hat q - \omega)
\nonumber\\
C_{5,3}^{\alpha,\mu\nu} (v, q,\omega_1,\omega_2) &=&  \frac{\pi}{2} g_\perp^{\mu\nu}
\frac{1}{m_b-\bar n \cdot q }
{\rm Tr } \left( s^\alpha\Gamma \nslash \overline{\Gamma} \right)\\
&&\qquad\times\left(\frac{\delta(1 - n \cdot \hat q - \omega_1) - \delta(1 - n \cdot \hat q -
\omega_2)}
            {\omega_1 - \omega_2} \right)
\nonumber  \\  \nonumber
C_{5,4}^{\alpha,\mu\nu} (n \cdot q,\omega_1,\omega_2)&=& (-i) \frac{\pi}{2}
\frac{1}{m_b-\bar n \cdot q }
{\rm Tr } \left( s^\alpha \Gamma \nslash (-i \sigma_{\!\!\perp}^{\mu\nu})
\overline{\Gamma}  \right) \nonumber\\
&&\qquad\times
\left(\frac{\delta(1 - n \cdot \hat q - \omega_1) - \delta(1 - n \cdot \hat q - \omega_2)}
            {\omega_1 - \omega_2} \right)
\nonumber \\ \nonumber
C_{5,T}^\alpha (n \cdot q,\omega)   &=&  C_{5,0}^\alpha (n \cdot q,\omega)\,.
\end{eqnarray}

\section{Matrix Elements}

Matrix elements of the subleading operators (\ref{nonlocalfteven}), (\ref{nonlocalftodd}) and (\ref{T-ops}) give rise to new, subleading structure functions. Writing
the most general Ansatz consistent with the symmetries and the equation of motion
$(i v\cdot D) \,h = 0$, we find that only the following matrix elements are
non-vanishing
\begin{eqnarray}
   \langle B(v) | O_1^\mu(\omega) | B(v) \rangle &=& 2 m_B \, g_1
(\omega) (v^\mu - n^\mu) \nonumber \\ \langle B(v) |
O_3^{\mu\nu}(\omega_1,\omega_2) | B(v) \rangle &=& 2 m_B \, g_2
(\omega_1, \omega_2) g_\perp^{\mu\nu} \nonumber \\ \langle B(v) |
P_{2,\alpha}^\mu(\omega) | B(v) \rangle &=& 2 m_B\, h_1(\omega)
\varepsilon_{\perp, \alpha}^\mu \\ \langle B(v) |
P_{4,\alpha}^{\mu\nu}(\omega_1,\omega_2) | B(v) \rangle &=& 2 m_B\, h_2
(\omega_1, \omega_2) \varepsilon_{\rho\sigma\alpha\beta} \,g_\perp^{\mu\rho}
g_\perp^{\nu\sigma}  v^\beta
\nonumber \\ \langle B(v) | O_T (\omega) | B(v) \rangle &=& 2m_B \,
t(\omega) \, ,\nonumber
\end{eqnarray}
where we define
\begin{equation}
\varepsilon_\perp^{\mu \nu} =
\varepsilon^{\mu \nu \alpha \beta} v_\alpha n_\beta\,,
\end{equation}
and $\varepsilon^{0123} = 1$.

The matrix element of $O_4^{\mu \nu} (\omega_1,\omega_2)$ between $B$ mesons
vanishes since no antisymmetric, parity even object can be constructed which is
perpendicular  to both $v$ and $n$. Similarly, the matrix element of
$P_{3,\alpha}^{\mu \nu} (\omega_1,\omega_2)$ vanishes, since there is no parity
odd, symmetric object perpendicular to $v$ and $n$. Due to the equations of
motion, the matrix element of $O_2^{\mu} (\omega)$ must be proportional to
$(v^\mu - n^\mu)$:
\begin{equation}
\langle B(v)|O_2^\mu(\omega) B(v)\rangle=a (v^\mu-n^\mu)\,.
\end{equation}
Contracting with $n^\mu$ we find
\begin{equation}
a=\langle B(v)|n \cdot O_2  (\omega) |B(v)\rangle =
\langle B(v)|
i\bar h_v \big[i n \cdot D ,\delta(in \cdot \hat D + \omega)\big] h_v |
B(v)\rangle = 0\,,
\end{equation}
(where the last equality is due to the delta function) and so the matrix element
vanishes. The matrix element of $P_{1,\alpha}^{\mu} (\omega)$ vanishes since all
of its moments with respect to $\omega$ vanish. Finally, the matrix element of
$P_{T,\alpha} (\omega)$ vanishes due to parity.

There is additional information on the remaining new functions. Starting with
$g_1(\omega)$, we find
\begin{eqnarray}
2 m_B g_1(\omega) &=& n_\mu \langle B(v) |  O_1^\mu (\omega) | B(v) \rangle
 = \langle B(v) |
\bar h_v \left\{i n \cdot D ,\delta(in \cdot \hat  D + \omega)\right\} h_v
| B(v) \rangle \nonumber  \\
&=& - 2 (m_b \,\omega) \langle B(v) |
\bar h_v \delta(i\hat D\cdot n+ \omega) h_v | B(v) \rangle
= -4 m_B (m_b \,\omega)\, f(\omega)\,.
\end{eqnarray}
Thus, $g_1(\omega)$ is determined by the leading order structure function,
$g_1(\omega) = -2(m_b \,\omega)\, f(\omega)$. Some information can also be
obtained on moments of the function $g_2(\omega_1, \omega_2)$:
\begin{eqnarray}
g_2^{(m,n)} &=& (-1)^{m+n} \int\! d\omega_1 d\omega_2 \omega_1^n \omega_2^m
g_2(\omega_1,\omega_2) \\
&=&  \frac{1}{2m_B}
\langle B(v) | \bar h_v (in \cdot D)^m (iD_\perp)^2 (in \cdot D)^n h_v | B(v) \rangle \,,
\nonumber
\end{eqnarray}
leading to
\begin{eqnarray}
g_2^{(0,0)} &=& \frac{2\lambda_1}{3}\,,\nonumber\\
g_2^{(m,0)} &=& g_2^{(0,n)} = 0\,, \qquad m,n \not= 0\,,
\end{eqnarray}
where the last equality arises because of the constraints from the equations of
motion \cite{tm94}
\begin{equation}
\langle B|\bar h_v(iD_\alpha)(i D_{\nu_1})\dots (i D_{\nu_n})(iD_{\beta}) h_v|B
\rangle=(g_{\alpha\beta}-v_{\alpha}v_{\beta}) {\cal A}_{\nu_1\dots\nu_n}.
\end{equation}
We can also obtain information on the parity odd operators. The function
$h_1(\omega)$ is a genuine new non-perturbative function which introduces spin
dependent effects. The first three moments of this function are given by
\begin{eqnarray}
\int\! d\omega \, h_1(\omega) = 0\,; \qquad
\int\! d \omega \, \omega \, h_1(\omega)  =   - \frac{\lambda_2}{m_b}\,; \qquad
\int\! d \omega \, \omega^2 h_1(\omega) =  \frac{\rho_2}{m_b^2}\, \,
\end{eqnarray}
where
\begin{eqnarray}
{1\over 2 m_B}\langle B|\bar h_v(iD_\alpha)(iD_\beta) s_\lambda h_v|B\rangle&\equiv&
{1\over 2} i\epsilon_{\nu\alpha\beta\lambda}v^\nu\lambda_2\\
{1\over 2 m_B}\langle B|\bar h_v(iD_\alpha)(iD_\mu)(iD_\beta) s_\lambda h_v|B
\rangle&\equiv&{1\over 2} i\epsilon_{\nu\alpha\beta\lambda}v^\nu v_\mu\rho_2.
\end{eqnarray}
The function $h_2(\omega_1, \omega_2)$ also introduces spin dependent effects
with the first few moments of the function given by
\begin{eqnarray}
h_2^{(0,0)} =  \lambda_2\,; \qquad h_2^{(1,0)} = 0 = h_2^{(0,1)}\,.
\end{eqnarray}
Finally we have to consider the function $t(\omega)$. Since $1/m_b$ terms are
absent in the total rate, we have
\begin{eqnarray}
\int\! d\omega \,  t(\omega) = 0  \,.
\end{eqnarray}
Furthermore, the first moment of $t(\omega)$ is related to $\lambda_1$
and $\lambda_2$
\begin{eqnarray}
2  m_B \int\! d \omega \, \omega \, t(\omega) &=& - i \int\! d^4 x \,
\langle B | T \left\{\big[\bar h_v (in\cdot \hat{D}) h_v\big](0)
           \,\, {\cal O}_{1/m} (x)\right\} | B \rangle
\\ \nonumber
&=&  2m_B \frac{\lambda_1 + 3 \lambda_2}{m_b} \,,
\end{eqnarray}
while the second moment introduces one new parameter
\begin{eqnarray}
\int\! d \omega \, \omega^2  \, t(\omega) = \frac{\tau}{m_b^2} \,,
\end{eqnarray}
where $3 \tau = - 2 ({\cal T}_1 + 3 {\cal T}_2)$ and ${\cal T}_i$ are the
parameters used in \cite{GK,Bauer}.

Combining these results with the known leading twist contribution (\ref{leadexp}) leads to
\begin{eqnarray} \label{fexp}
f(\omega) &=& \delta(\omega)
              - \frac{\lambda_1}{6m_b^2} \delta '' (\omega) -
                \frac{\rho_1}{18m_b^3} \delta ''' (\omega) + \cdots \nonumber\\
\omega f(\omega) &=&  \frac{\lambda_1}{3m_b^2} \delta'(\omega) +
      \frac{\rho_1}{6m_b^3} \delta '' (\omega) + \cdots  \nonumber\\
h_1(\omega) &=& \frac{\lambda_2}{m_b} \delta'(\omega)
               +\frac{\rho_2}{2m_b^2} \delta''(\omega) + \cdots  \\
g_2 (\omega_1,\omega_2) &=& \frac{2\lambda_1}{3}
                            \delta(\omega_1) \delta(\omega_2) + \cdots  \nonumber\\
h_2 (\omega_1,\omega_2) &=& \lambda_2
                            \delta(\omega_1) \delta(\omega_2) + \cdots
 \nonumber\\
t(\omega) &=& - \frac{\lambda_1 + 3 \lambda_2}{m_b} \delta ' (\omega)
              + \frac{\tau}{2m_b^2} \delta '' (\omega) + \cdots \,,\nonumber
\end{eqnarray}
where we have used the relations
$$
\xi \delta '' (\xi) = - 2 \delta ' (\xi);\quad
 \xi \delta ''' (\xi) = -3 \delta '' (\xi)\,,
$$
which are true when integrated against a function which is non-singular as $\xi \to 0$. 

\section{Application to $\bar B\rightarrow X_s\gamma$}

The decay $\bar B \to X_s \gamma$ is described by the effective Hamiltonian
\begin{equation}
H_{\rm eff} = - \frac{G_F^2}{\sqrt{2}} V_{tb} V_{ts}^* C_7(\mu) {\cal O}_7+\dots\,,
\end{equation}
where
\begin{equation}
{\cal O}_7 = \frac{e}{32 \pi^2} m_b \bar{s}
             \sigma_{\mu \nu} (1+\gamma_5) b \, F^{\mu \nu} \,,
\end{equation}
and the dots denote additional operators which we shall neglect for the purposes
of this discussion. This effective Hamiltonian leads to the Dirac structure
\begin{equation}
\Gamma = -i A \, \sigma^{\mu\nu}\varepsilon_\mu q_\nu (1+\gamma_5)
\quad \mbox{ with } \quad A = \frac{G_F^2}{\sqrt{2}} V_{tb} V_{ts}^* C_7(\mu)  \frac{e}{16 \pi^2} m_b\,.
\end{equation}
The kinematics of this decay are particularly simple since $q^2=0$,
\begin{eqnarray}
q^\mu = E_\gamma \bn^\mu \equiv x \frac{m_b}{2} \bn^\mu \,
\end{eqnarray}
and so the large and small kinematical factors defined in section
\ref{kinematics} are
\begin{eqnarray}
m_b-\bn \cdot q = m_b\,, \qquad m_b-n \cdot q = m_b(1-x)\,.
\end{eqnarray}

Computing the matching coefficients, we find for the rate
\begin{eqnarray} \label{resultbsg}
\frac{d \Gamma}{dx}  &=& \Gamma_0
\left\{ f(1-x) - 2(1-x) f(1-x)  + \frac{1}{2m_b}t(1-x) +\frac{1}{m_b} h_1(1-x) \right.
\nonumber \\ && \qquad \qquad \left.
-  \frac{1}{m_b^2}
\Big[ G_2(1-x) - H_2 (1-x)\Big] \right\} \,,
\end{eqnarray}
where
\begin{equation}
\Gamma_0 = \frac{G_F^2 \alpha |V_{ts} V_{tb}^*|^2 |C_7 (m_b)|^2 }{32 \pi^4} m_b^5\,,
\end{equation}
and
\begin{eqnarray} \label{Dfunctions}
G_2 (1-x) &=& \int\! d\omega_1 \, d\omega_2 \, g_2(\omega_1, \omega_2)
\left[\frac{\delta(1 - x - \omega_1) - \delta(1 - x - \omega_2)}{\omega_1 -
\omega_2} \right]  \\ H_2 (1-x) &=& \int\! d\omega_1 \, d\omega_2 \,
h_2(\omega_1, \omega_2) \left[\frac{\delta(1 - x - \omega_1) - \delta(1 - x -
\omega_2)}{\omega_1 - \omega_2} \right] \,.
\end{eqnarray}

This result can be compared with previous work \cite{Bauer}, where the $1/m_b$
expansion has been discussed. This expansion is recovered by performing a moment
expansion of (\ref{resultbsg}). Using (\ref{fexp}) we find
\begin{eqnarray}
G_2(1-x) &=& -\frac{2\lambda_1}{3} \delta'(1-x) + \cdots\nonumber\\
H_2(1-x) &=& -\lambda_2 \delta'(1-x) + \cdots\,,
\end{eqnarray}
leading to
\begin{eqnarray}
\frac{d \Gamma}{dx} &=& \Gamma_0 \left[ \delta(1-x) -
\frac{\lambda_1+3\lambda_2}{2m_b^2}
\,\delta'(1-x) \right.
\\ \nonumber
&& \left. \qquad \qquad
- \left( \frac{\lambda_1}{6m_b^2}
+ \frac{2 \rho_1 - 3
\rho_2 + {\cal T}_1 + 3{\cal T}_2}{6m_b^3} \right) \delta''(1-x)
 - \frac{\rho_1}{18m_b^3} \delta'''(1-x) \right] \,,
\end{eqnarray}
in agreement with \cite{Bauer}.
{}From the matching coefficients (\ref{coeffeven}) we see that the Wilson
coefficients $C_T$ is identical to the leading order coefficient. This implies
that for any current mediating a heavy-to-light decay the subleading structure
functions $t(\omega)$ always arises in the same linear combination with the
leading order function $f(\omega)$. Thus, we can always combine the functions
$f(\omega)$ and $t(\omega)$ into a new universal function which is  defined by
\begin{equation}
F(\omega) = f(\omega) + {1\over 2 m_b} t(\omega)\,.
\end{equation}
This new structure function has the moments
\begin{eqnarray}
\int\! d\omega F(\omega) = 1\,;
\qquad\int\! d\omega \, \omega \, F(\omega) = \frac{
  \lambda_1 + 3 \lambda_2}{2m_b^2}\,; \qquad\int\! d\omega \, \omega^2 F(\omega) =
  -\frac{\lambda_1}{3m_b^2} + \frac{\tau}{2m_b^3}\,.
\end{eqnarray}
There are three new subleading twist structure functions; the spin independent
function $G_2(\omega)$, as well as  $h_1(\omega)$ and  $H_2(\omega)$, which are
sensitive to the heavy quark spin. Thus there are in total four functions
parametrizing the heavy-to-light decays to subleading twist, since we may replace
\begin{equation}
\omega f(\omega) = \omega \,F(\omega)+\dots\,,
\end{equation}
where the dots denote higher twist terms.

\subsection{A Simple Model}

To get some insight into the size of the effect of these new functions we will
use a simple model for these functions which incorporates the information we have
on their moments.  The following model has been proposed\cite{mn} for the leading
twist distribution function
\begin{eqnarray}
f_\model(\omega) = \frac{32 m_b}{\pi^2\bar\Lambda} (1-y)^2
            \exp \left[ -\frac{4}{\pi} (1-y)^2 \right]\Theta(1-y)\,; \qquad
        y = - \frac{m_b \,\omega}{\bar\Lambda}  \,,
\end{eqnarray}
where $\bar\Lambda$ is the only free parameter.  This model therefore assumes a
simple correlation between all higher moments of $f(\omega)$.

All of the new subleading twist functions have vanishing zeroth moments. In order to construct a
model for these subleading functions we use the derivative of the leading twist
function. Normalizing this derivative to match
the known first moments of the subleading functions, we obtain
\begin{eqnarray}\label{Ffunct}
F_\model(\omega) &=& f_\model(\omega) - \frac{\lambda_1 + 3 \lambda_2}{2 m_b^2}
            f_\model'(\omega)  \nonumber\\
G_{2\,\model}(\omega) &=& - \frac{2 \lambda_1}{3} f_\model'(\omega)  \,.
\end{eqnarray}
For the decay $\bar B \to X_s \gamma$ the two functions $h_1(\omega)$ and
$H_2(\omega)$ enter in the combination
\begin{eqnarray}
\Delta_h(\omega) = h_1(\omega) + H_2(\omega)/m_b\,.
\end{eqnarray}
The zeroth and first moments vanish, while it has a non-vanishing second
moment
\begin{eqnarray} \label{delmom}
\int\! d\omega \, \omega^2 \Delta_h(\omega) = \frac{\rho_2}{m_b^2}\,.
\end{eqnarray}
In our simple approach to modeling the subleading functions,
we would obtain $\Delta_{h, \model}(\omega) \equiv 0$, since the first moments of
$m_b h_1 (\omega)$ and $H_2 (\omega)$ coincide.  To use the information on the second
moment (\ref{delmom}), we instead model $\Delta_h(\omega)$ by the second derivative of
$f_\model$:
\begin{equation}
\Delta_{h,\model}(\omega) = \frac{\rho_2}{2m_b^2} f_\model''(\omega)\,.
\end{equation}
This leads finally to our model for the differential
decay spectrum of the decay $\bar B \to X_s \gamma$:
\begin{eqnarray}\label{diffmodel}
\frac{d \Gamma_\model}{dx} &=& \Gamma_0 \left[(2x-1)F_\model(1-x) +
  \frac{2 \lambda_1}{3m_b^2} f_\model'(1-x)+
  \frac{\rho_2}{2m_b^3} f_\model''(1-x) \right] \, \\
&=& \Gamma_0 \left\{ (2x-1)
\left[ f_\model(1-x) - \frac{\lambda_1 + 3 \lambda_2}{2 m_b^2}
            f_\model'(1-x)\right]+ \frac{2 \lambda_1}{3m_b^2}
f_\model'(1-x)\right.\nonumber\\ &&\qquad
+\left. \frac{\rho_2}{2m_b^3} f_\model''(1-x) \right\} \, .\nonumber
\end{eqnarray}

\begin{figure}
\centerline{\epsfysize=7truecm \epsfbox{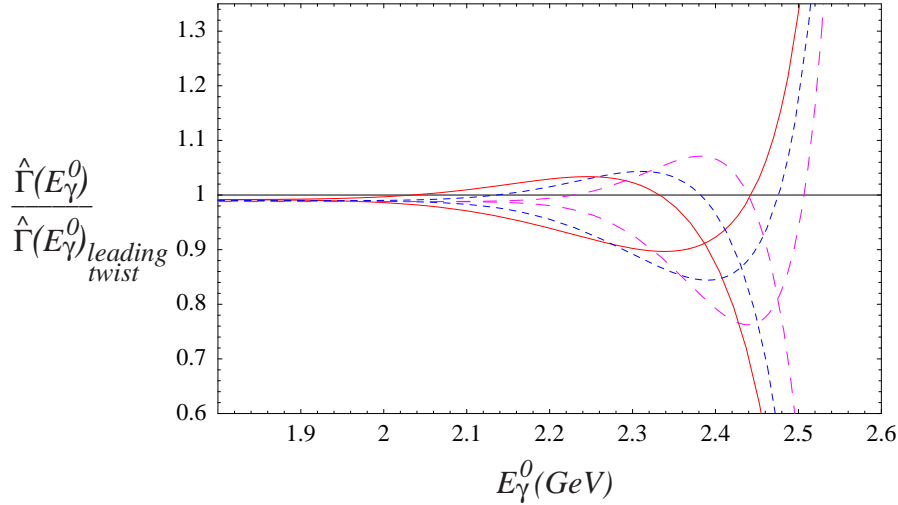} }
\caption[]{Partially integrated Rate
\protect{(\ref{partial})}, normalized to the leading twist result, using the 
simple model
given in \protect{(\ref{diffmodel})}.
The solid, short-dashed and long-dashed lines correspond to $\bar\Lambda=570$
MeV, 470 MeV and 370 MeV, respectively.  The lines which rise at the endpoint
correspond to $\rho_2=(500\ {\rm MeV})^3$, while those that go down correspond
to $\rho_2=-(500\ {\rm MeV})^3$.  The values of $\lambda_1$ has been chosen to
reproduce the second moment of the leading order structure function, $\lambda_1
=-0.53\ \bar\Lambda^2$.}
\label{partialplot}
\end{figure}

We can now use this spectrum to analyze the effect of the subleading twist
contributions to the partially integrated decay rate
\begin{eqnarray} \label{partial}
\hat\Gamma(E^0_\gamma) = \frac{1}{\Gamma_0} \int_{x_0}^{x_{\rm max}} \frac{d
  \Gamma}{d x} dx \, ,\quad  x_{\rm max} = \frac{m_B}{m_b}
\mbox{ and  } x_0 = \frac{2E^0_\gamma}{m_b} \,,
\end{eqnarray}
with $E^0_\gamma$ being a lower cut on the photon energy. The effects of the
subleading shape functions are shown in Fig.~\ref{partialplot}, in which we plot
the ratio of the partially integrated rate with and without the subleading twist contributions as a function of the photon
energy cut for various values of the parameters $\bar\Lambda$ and $\rho_2$. With this
simple model, the curves on this plot should only be taken as an
estimate of the size of the corrections in different kinematic regions. Recall
that we expect that for a photon cut $E^0_\gamma\mst{<}{\sim}m_B/2-\lqcd \sim
2.1\ \GeV$ the usual OPE should hold, while for $m_B/2-\lqcd\sim 2.1\
\GeV\mst{<}{\sim}E^0_\gamma\mst{<}{\sim} m_B/2-\lqcd^2/m_B\sim 2.5\ \GeV$ the
twist expansion presented in this paper is appropriate, with subleading twist 
corrections naively of order $\lqcd/m_b\sim 10\%$.  For $E^0_\gamma\mst{>}{\sim}
m_B/2-\lqcd^2/m_B\sim 2.5\ \GeV$ we are in the resonance region, and the twist
expansion is expected to break down.  From the figure we see that these expectations
are borne out in our simple model:  below $\sim 2.1\ \GeV$ the corrections to the
leading twist are small, since they are only corrections of order $\lqcd^2/m_b^2$ to
the leading rate.  Below $E^0_\gamma\sim 2.4\ \GeV$, the corrections are less than
or of order $\pm$20\%, while the twist expansion starts to break down above this cutoff.

\section{Conclusions}

It has been known for some time that in inclusive heavy hadron decays the naive
short distance expansion must be replaced by a twist expansion if the phase space
is restricted to a region of large energy, low invariant mass final hadronic
states. The leading term, parametrized by the the light-cone distribution
function of the heavy quark in the hadron, is well investigated by now, but
subleading terms of this expansion have not been previous studied. In the present
paper we have identified the non-local operators appearing at subleading order in
the twist expansion. The tree level matching to these operators has been computed
for a general bottom hadron decay and the matrix elements of the subleading
operators have been parametrized for $B$ meson decay.

We found that for any inclusive $B$ meson decay four independent subleading
distribution functions are needed. We worked out the case for $\bar B\to X_s
\gamma$ in detail.  Using a simple model for the leading and the subleading
distribution functions we studied the effects of the subleading terms on the
photon energy spectrum. We found that they had the expected behavior:  in the
region where the local OPE is appropriate, these corrections were negligible,
whereas in the region where the twist expansion was appropriate, they were of
order 10-20\%, depending on the parameters of the model.

Since the leading distribution function is not known, much less to 10-20\%
accuracy, these results are of limited utility for $\bar B\rightarrow X_s\gamma$
decays (although they do indicate the region where the twist expansion breaks
down). However, there are certain relations between the charged lepton energy
spectrum in $\bar B\rightarrow X_u\ell\bar\nu$ and the photon spectrum in $\bar
B\rightarrow X_s\gamma$ for which the leading distribution function drops
out\cite{mn,sudakovs}.  In this case, even a model of the subleading distribution
functions will provide a useful estimate of the theoretical uncertainty in these
relations, and the resulting extraction of $|V_{ub}|$.  This investigation is in
progress \cite{blmprog}.

\section*{Acknowledgments}

We thank Iain Stewart for comments on the manuscript. This work was
supported by the Department of Energy under grants DOE-FG03-97ER40546 and
DOE-ER-40682-143 and the Natural Sciences and Engineering Research Council of
Canada, and by the DFG Graduiertenkolleg ``Elementarteilchenphysik an
Beschleunigern'', from the DFG Forschergruppe ``Quantenfeldtheorie,
Computeralgebra und Monte Carlo Simulationen'' and from the Ministerium f\"ur
Bildung und Forschung bmb+f. CWB and ML gratefully acknowlege the hospitality of
the CERN TH division, where this work was completed.

\end{document}